# Cosmic Ray Accelerators in the Large Magellanic Cloud

Yousaf M. Butt[1]

[1]Harvard-Smithsonian Center for Astrophysics, Cambridge, Massachusetts 02138, USA. ybutt@cfa.harvard.edu

## ABSTRACT

I point out a correlation between gamma-ray emissivity and the historical star formation rate in the Large Magellanic Cloud ~12.5 Myr ago. This correlation bolsters the view that CRs in the LMC are accelerated by conglomerations of supernova remnants – i.e. superbubbles and supergiant shells.

## INTRODUCTION

The origin of Galactic cosmic rays (GCRs) is a century old enigma. Superbubbles (SBs) have been proposed as GCR acceleration sites due to their great power, scale and duration (e.g., Bykov 2001; Parizot et al., 2004; Butt 2009). They are powered by the fast stellar winds and multiple powerful supernova explosions of massive stars in dense stellar clusters and associations. Although visible in the radio and X-ray bands, Galactic superbubbles have not yet been positively detected in $\gamma$-rays even though the importance of their role in GCR acceleration has been inferred from GCR composition (Higdon and Ligenfelter, 2005). It is, as yet, unclear what the precise acceleration mechanism within

SBs is, and thus it is important to have some observational grounding of the theoretical models (e.g. Ferrand & Marcowith, 2010).

Recent evidence from very high energy observations of external Galaxies with high star formation rates (SFRs) – i.e. in which almost all supernovae would be expected to occur within SBs created by nearby overlapping SNRs – bolster the conjecture that SBs, and similar structures such as supergiant shells (SGSs) [Meaburn, 1980], accelerate CRs. The M82 starburst galaxy has been detected by the VERITAS collaboration [ Acciari et al., 2009] and NGC 253 by the HESS observatory [ Acero et al., 2009 ] in the TeV gamma-ray band. These detections appear to show that CRs are accelerated by conglomerations of SNRs in such Galaxies, although it cannot yet be ruled out that other objects (e.g. pulsars) may be responsible for the detected gamma-ray emission: indeed, most energetic objects capable of accelerating CRs are expected to reside in high SFR regions.

In this note, I provide some further evidence that the SBs and SGSs associated with star forming regions in the neighboring Large Magellanic Cloud (LMC) galaxy are likely responsible for accelerating CRs there, as previously suggested by Butt and Bykov (2008) based on simple energetics arguments.

## Gamma-Ray Correlation with Star Formation History

Recently, the Fermi orbiting gamma-ray observatory detected γ-ray emission from the LMC, and provided the first spatially-resolved view of γ-rays from a nearly face-on external galaxy (Abdo et al., 2010). The LMC's distance of 50 kpc and low inclination angle make it possible to compare the distribution of γ-ray emission with the underlying stellar population and interstellar structures, for a critical examination of the possible sites of CR acceleration.

Fermi LAT observations of the LMC has found the brightest γ-ray emission centered near the 30 Dor giant HII region, with fainter γ-ray emission also detected in the northern part of the LMC. The γ-ray emission detected by Fermi shows little correlation with the total column density of the interstellar gas and the γ-ray emission appears to be coincident with massive star forming regions. These findings indicate that CRs in the LMC are likely accelerated in massive star forming regions and that the diffusion length of GeV-range CR protons is relatively short (Abdo et al. 2010).

Although Abdo et al (2010) find a generally good correlation of the gamma-ray emissivity with the 30 Dor star forming region in the central region of the LMC, there is an "orphan" region of fainter γ-ray emission towards the north with no similar concrete counterpart yet identified. In order to better understand the origin of the gamma-ray emission (and thus the CRs) I compare the distribution of γ-ray emission with the underlying stellar and interstellar components.

A detailed study of the star formation history of the LMC is provided by Harris & Zaritsky (2009). In fact, the integrated >100 MeV γ-ray emissivity map of the LMC correlates very well with the star formation rates 12.5 Myr ago in the LMC as deduced by

Harris & Zaritsky (2009). Importantly, this is true not only for the $\gamma$-ray peak coincident with 30Dor (as already noted by Abdo et al., 2010) but also for the fainter northern $\gamma$-ray emission – as well as for the extension to the west (Fig. 1). Since the progenitors of supernovae in the LMC have a lifetime ranging from a few to ~15 Myr, the spatial coincidence of the $\gamma$-ray emissivity with the sites of 12.5 Myr old star formation indicates that these conglomerations of supernovae (i.e. correlated in time and space), play a major role in the acceleration of CRs in the LMC.

An examination of the H$\alpha$ image[1] and HI column density map (Kim et al, 2003) of the LMC also reveals superbubbles and supergiant shells in regions where the star formation rate was high within the last ~12 Myr (Fig 1). Therefore, the $\gamma$-ray emission is also well correlated with superbubbles and supergiant shells.

## Supporting Evidence for CR acceleration by SBs in the LMC

Further circumstantial evidence for the SB acceleration of CRs in the LMC includes the fact that the observed thermal and kinetic energies of several SBs there are significantly lower than the stellar and supernova energy input. e.g. Observations of the SB "DEM L192" show that it contains only about one-third the energy injected by its constituent stars via fast stellar winds and supernovae (Cooper et al., 2004), most likely implying that the "missing" energy has gone into accelerating CRs (Butt and Bykov 2008). The presence of diffuse nonthermal X-ray emission (30 Dor: Bamba et al. 2004; DEM L192: Cooper et al. 2004; N11: Maddox et al., 2009) further bolsters this view.

---

1 http://www.ctio.noao.edu/~mcels/

**Conclusions**

The above discussion supports the conjecture that the collective, interacting SNR shocks within SBs and SGSs (produced by massive stars formed in the last ~15Myr) have likely accelerated the CRs in the LMC that are responsible for the >100 MeV $\gamma$-rays detected by Fermi.

Further study of the $\gamma$-ray emissivity of the LMC, complemented by a detailed knowledge of star formation history and interstellar gas structure, will help localize where precisely CRs in the LMC are (and were) accelerated and how CRs diffuse into the interstellar medium. Deeper $\gamma$-ray data from Fermi and other observatories will also reveal whether the outlying regions of star formation 12.5 Myrs ago (Fig. 1) are eventually confirmed as $\gamma$-ray emitters.

Understanding the process of CR acceleration in the LMC will complement the theoretical investigations of the CR acceleration process at work in SBs (eg. Ferrand & Marcowith, 2009) and will allow us to construct better models, both for the LMC and for our own Galaxy. For example, does CR acceleration in SBs take place via a bunch of isolated SNRs or do SBs really act as one large, coherent, long-lasting, accelerator?

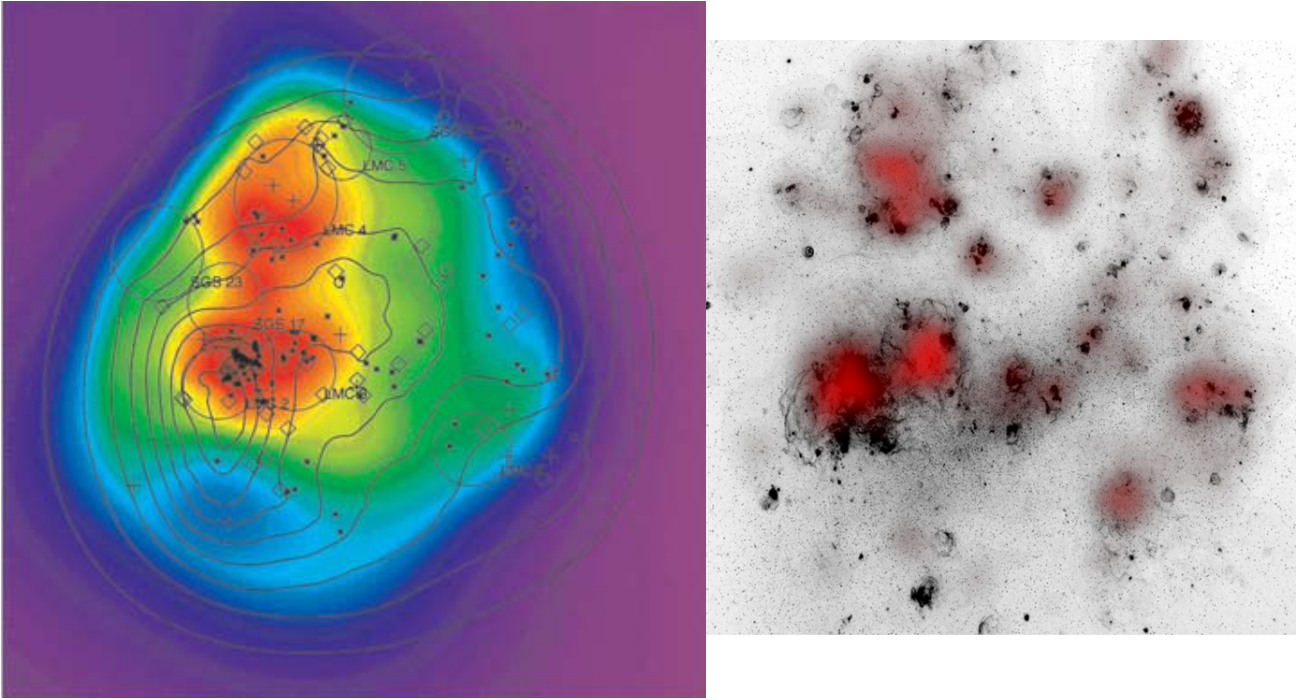

Figure 1. ( Left) Luminosity map of the Large Magellanic Cloud (LMC) from Fermi γ-ray telescope data for γ-rays with energy above 100 MeV. Contour lines indicate density of hydrogen gas and colors indicate local γ-ray emission per hydrogen gas atom. From Abdo et al., 2010  (Right) Correlation of the recent (Age < 12.5 Myr) star formation activity in the LMC based on the analysis of Harris & Zaritsky (2009), with the Hα image of the LMC from the MCELS[1]. From Harris & Zaritsky (2009). *The two images have been (coarsely) rescaled to match by the author*.

## Acknowledgments

I acknowledge partial support from a NASA Long Term Space Astrophysics Grant, and would also like to thank You-Hua Chu and Robert Gruendl for insightful discussions.